\documentclass[10pt,twocolumn,letterpaper]{article}
\usepackage{btas}
\usepackage{times}
\usepackage{epsfig}
\usepackage{graphicx}
\usepackage{amsmath}
\usepackage{amssymb}
\usepackage{multirow}
\usepackage{subfigure}
\usepackage[none]{hyphenat}
\usepackage{epsfig}
\usepackage{epstopdf}
\usepackage[hyphens]{url}
\usepackage[numbers,sort&compress]{natbib}
\usepackage[margin=0.8in,bottom=0.9in,top=1in]{geometry}
\btasfinalcopy 

\ifbtasfinal\fi

\begin{document}

\title{Continuous User Authentication via Unlabeled Phone Movement Patterns}
\author{Rajesh Kumar\\
Syracuse University\\
NY, USA\\
{\tt\small rkuma102@syr.edu}
\and
Partha Pratim Kundu\\
Syracuse University\\
NY, USA\\
{\tt\small ppkundu@syr.edu}
\and
Diksha Shukla\\
Syracuse University\\
NY, USA\\
{\tt\small dshukla@syr.edu}
\and
Vir V. Phoha\\
Syracuse University\\
NY, USA\\
{\tt\small vvphoha@syr.edu}}
\maketitle
\thispagestyle{empty}

\begin{abstract}
\par In this paper, we propose a novel continuous authentication system for smartphone users. The proposed system entirely relies on unlabeled phone movement patterns collected through smartphone accelerometer. The data was collected in a completely unconstrained environment over five to twelve days. The contexts of phone usage were identified using k-means clustering. Multiple profiles, one for each context, were created for every user. Five machine learning algorithms were employed for classification of genuine and impostors. The performance of the system was evaluated over a diverse population of $57$ users. The mean equal error rates achieved by Logistic Regression, Neural Network, kNN, SVM, and Random Forest were $13.7\%$, $13.5\%$, $12.1\%$, $10.7\%$, and $5.6\%$ respectively. A series of statistical tests were conducted to compare the performance of the classifiers. The suitability of the proposed system for different types of users was also investigated using the failure to enroll policy.
\footnote{Copyright (c) 2017 IEEE\\
Personal use of this material is permitted. Permission from IEEE must be obtained for all other uses, in any current or future media, including reprinting/republishing this material for advertising or promotional purposes, creating new collective works, for resale or redistribution to servers or lists, or reuse of any copyrighted component of this work in other works.
To appear in IEEE International Joint Conference on Biometrics (IJCB 2017), Denver, Colorado}
\end{abstract}
\vspace{-0.2in}
\section{Introduction}
\label{Introduction}
\par The PIN, password, and pattern based authentication systems have several drawbacks as they need to be remembered; are not user-friendly anymore--as the overall authentication process is time consuming; offer only entry-point security; and are susceptible to video-based side channel, shoulder surfing, and social engineering attacks \cite{ShuklaCCS,PatternBreak}. The fingerprint, face, and iris based recognition systems do address the first two drawbacks, however, they are also susceptible to spoof, and social engineering (intoxication) attacks, as well as, not very suitable for continuous verification \cite{FingerPrintScanner}. The above-mentioned pitfalls of the existing and extensively used authentication systems are some of the key reasons behind the rapidly evolution of behavioral-biometrics continuous authentication systems since the last few years.
\par Behavioral footprints such as typing, swiping, walking, arm movements while walking, hand kinematic synergies and their neural representations, and possible combination of these have demonstrated the potential, and promise in user authentication \cite{PhoneTyping4,PhoneSwiping3,Abena,ArmMovement,TypingSwipingFusion,HMOG,HandNeuralSynergy,SmartWatchWisdom}. However, several challenges still exist and have not been addressed. For example, the availability of these modalities throughout the user interaction with the smartphone; collection of labeled data for enrollment and verification process under different operating conditions (phone-usage contexts), and practicality of these systems under realistic scenarios.
\par Moreover, the context of phone usage varies from user to user significantly. For example, a smartphone user can swipe, type, talk while sitting, standing, walking, in an elevator, in a moving bus, train, or in a car. For each of these contexts, the biometric footprints may vary significantly. So building authentication systems without taking the context into consideration would result in poor authentication decisions. Additionally, the availability of these footprints may vary across different applications. In practice, the existing individual modality based systems could have limited application, such as to secure activities on a targeted application. Almost every study has assumed the phone-usage context, and/or the availability of labeled samples to guide the authentication process, and have maintained constrained data collection environments \cite{ChungXu, Abena, TypingSwipingFusion, HMOG}.
\par In reality, the process of labeling contexts is quite intrusive and unlikely to be implemented by industry and/or accepted by common smartphone users. Hence, both of the assumptions i.e. context is known, and availability of labeled samples are unrealistic. Some smartphone users may have specific usage contexts occurring with very high frequency during typical phone usage, while the other may never operate their smartphones in some of the contexts at all. Therefore, assuming a universal set of contexts, common across the user population may not be very helpful. Instead, developing models that could identify user-specific phone-usage contexts, and designing authentication models for each context could be a plausible solution.
\par This paper attempts to address the above-mentioned challenges by collecting phone movement patterns that are seamlessly available while the phone is in use or just in user's possession under a completely unconstrained environment (i.e. makes no attempt to label the data or assumes any context); applies semi-supervised learning algorithm to identify phone-usage contexts, and utilizes the predicted contexts to guide the authentication process. The main contribution of this paper is summarized below:
\begin{itemize}
\item Builds a dataset of continuous phone movement patterns collected from a diverse population of $57$ users over a period of $5$ to $12$ days under a completely unconstrained environment.
\item Presents a novel authentication method based only on phone movement patterns. The method identifies the phone-usage context automatically by using K-means clustering and Random Forest classifier. The enrollment and verification process was implemented by employing five distinct machine learning classifiers, namely, Logistic Regression, Neural Network, kNN, SVM, and Random Forest.
\item The performance of the authentication system was evaluated and reported in terms of Equal Error Rates (EERs). The performance of classification algorithms was compared using a series of statistical tests.
\item The suitability of the proposed system for different types of users was also investigated using Failure to Enroll policy. This investigation provided interesting insights into the usability of the proposed system.
\end{itemize}
\par The rest of this paper is organized as follows: Section \ref{RelatedWork} presents related work; Section \ref{DataCollectionPreprocessing} describes the data collection, preprocessing, and feature analysis; Section \ref{ExperimentalSetup} discusses design, implementation, training and testing processes of the proposed authentication system; Section \ref{ExperimentalResultsAndDiscussion} presents the experimental results and discussion; and Section \ref{ConclusionAndFutureWork} concludes our work.
\section{Related Work}
\label{RelatedWork}
Several behavioral footprints including swiping \cite{PhoneSwiping3}, typing \cite{PhoneTyping4}, walking \cite{Abena}, arm movements \cite{ArmMovement}, and their possible fusion \cite{TypingSwipingFusion, HMOG} have been studied in the past for authenticating individuals continuously \cite{VishalSurvey}. The phone movement pattern was studied by Kumar et al. \cite{TypingSwipingFusion}, Sitova et al. \cite{HMOG} and, Buriro et al. \cite{SwipeAndMovement} recently. Sitova et. \cite{HMOG} focused mainly on phone movement patterns while walking and sitting, whereas, Kumar et al. mainly studied phone movement patterns while typing or swiping. In a different study, Tang et al. \cite{ChungXu} studied the phone movement patterns under two different conditions, dynamic (walk, upstairs, and downstairs), and static (sit, stand, and lie). Murmuria et al. \cite{MurmuriaPoWer} studied power consumption, touch gestures, physical movement, and their combination for continuous authentication of smartphone users.

\par One of the common problems that the above studies have that their experiments have been carried out under a controlled or restricted environment. Another major problem that they have is the availability of the mentioned footprints thought the user activity on the device. Our work addresses these concerns by  (1) keeping the data collection environment completely unconstrained, and (2) capturing the phone movement patterns throughout the user interaction with the device.

\par In a similar attempt, Mahbub et al. \cite{Mahbub} collected data from multiple sensors including front camera, touch sensor, and location service under an unconstrained environment for continuous authentication of smartphones users. In another study, Mehbub et al. \cite{PATH} proposed methods to authenticate users based on their trace histories. They also present a challenging dataset that contains multiple sensor signals collected from $48$ volunteers on Nexus 5 phones over a period of $2$ months. The sensors include a front-facing camera, touchscreen, gyroscope, accelerometer, magnetometer, light sensor, GPS, Bluetooth, Wi-Fi, proximity sensor, temperature sensor and a pressure sensor. It would be interesting to apply our methods on the dataset presented by Mahbub et al. \cite{Mahbub} and explore the viability of all the modalities that have been captured for continuous authentication of smartphone users.
\section{Data Collection and Preprocessing}
\label{DataCollectionPreprocessing}
\subsection{Data Collection}
\par Following approval of the University's Institutional Review Board, we invited the faculty, staff, and students to participate in our study. The majority of the participants were university students, while the rest were university staff or faculty. All of them were regular smartphone users. The participating individuals were from different colleges, departments, and programs, including Engineering, Mathematics, Biomedical and from different countries including India, China, Nepal, Guyana, Uganda, USA, Iran, and Russia, resulting in a diverse sample. The participants were briefed about the level of engagement, type of data collection, battery consumption, our expectation, and the amount of compensation.

\par An Android application (App) was developed to capture phone movements patterns continuously as long as the phone was switched on. On its first startup, the App automatically activated a service in the background that was responsible to capture the acceleration of the phone seamlessly. The App was installed on the participant's phone instead providing them an experimental phone to ensure completely realistic and unconstrained operating environment.

\par However, there was at least one disadvantage of doing so was the varying sampling rate across the user population. The configuration for sampling rate was set to \textit{sensor\_delay\_normal} to avoid excessive power consumption by the participant's device. The rule of sampling rate was not obeyed by all of the devices as they were running different versions of the operating systems (Android $4.0$ or higher). The forty of the total $57$ users had sampling rate between $4$ to $10$, six users had between $10$ to $20$, whereas, remaining $11$ users had more than $20$ samples per second. The mean of sampling rate was $18.28$ whereas the median was $6.63$. To deal with the varying sampling rates, we used a fixed length of windows in terms of time instead of number of samples during feature extraction.

\par The participants were asked to come in at least five days or later, give the data, and collect their compensation of $\$30$. Several individuals registered to participate in our data collection study, about half of them never returned. We selected only those users who had at least five and up to twelve days of data for this study. The total number of users who followed the $5-12$ days criteria turned out to be $57$ out of $88$ who returned. The data was collected for several days to ensure that a full cycle of activities that the user undertake is captured.
\subsection{Exploratory Data Analysis}
\par For almost every user, a major segment of the data belonged to the unattended (but switched on) state of the phone. The unattended state of the phone in this context means the accelerometer records almost no acceleration. 
The direct removal of the segment of data that had lower acceleration than a threshold was distorting the signal. So, we used a window based scheme to identify and remove the segments belonging to the unattended state of the phone. For each user, the data was divided into segments of 2.5 seconds. For each segment, the medians of x, y, and z values were computed and compared with predefined thresholds. The segment was discarded using the following criteria:
if ($l_{x}<m_x<u_{x}$) \&\& ($l_{y}<m_y<u_{y}$) \&\& ($l_{z}<m_z<u_{z}$) then discarded else kept, where $m_x$, $m_y$, and $m_z$ were the medians of x, y, and z values, and $l_{x}$, $u_{x}$, $l_{y}$, $u_{y}$, $l_{z}$, and $u_{z}$, were lower and upper thresholds for series x, y, z respectively. The thresholds were computed by using the upper and lower envelope of the accelerometer readings that were collected by keeping the phone unattended for hours. The values of the thresholds were -$0.036$, $0.035$, $-0.02$, $0.06$, $-0.22$, and $-0.13$ respectively.
\par The data belonging to unattended phone state was discarded for two reasons. First, the phone movement pattern while the phone was unattended, was overlapping too much across the user population, so it could have injected only noise into the authentication system. Second, we believed that there is no need to authenticate the user when the phone is in the unattended state. The median filtering with a span of three data points was applied before extracting features.
\begin{table*}[htp]
\centering
\scriptsize
\caption{The list of features extracted from phone movement patterns recorded by accelerometer sensors built into smartphones.}
\vspace{0.1in}
\begin{tabular}{|l|l|l|l|l|}
\hline
Spectral entropy & Histogram (16 bins)   & Standard deviation & Interquartile Range $(Q_{3}-Q_{1})$ & Mean  \\ \hline
Bandpower & Dynamic Time Warping distance between pair of signals & Range (max-min)  & Peak magnitude to RMS Ratio & Energy  \\ \hline
Median frequency & Mutual information between pair of signals & Screen on/off & Correlation between pair of signals& \\ \hline
\end{tabular}
\label{FeatureList}
\end{table*}

\section{Experimental Setup}
\label{ExperimentalSetup}
\label{SystemDesignImplementationTesting}
\subsection{Characteristics of Phone Movement Pattern}
\par We hypothesized that the phone movement patterns recorded under different operational contexts could be used to build continuous authentication systems. The hypothesis was based on the fact that the phone movements are readily available, and easily collectible throughout the user interaction with the device, exerted by almost every user, could be distinctive if observed under specific contexts, do not change frequently, neither requires any user attention or intervention, and hardly imitable \cite{JainBiometric}. The initial design goals of the system aimed to be able to: run in the background; identify the context of the phone usage automatically; select the corresponding authentication model; and verify individual's authenticity at frequent intervals.

\par The phone movement patterns were not distinctive enough among the users as we expected in the first observation. However, when divided into several contexts through clustering, they turned out to be quite distinctive measurements among the users \cite{HMOG, TypingSwipingFusion}. So one of the biggest challenges that we faced was to automate the identification of contexts from the unlabeled phone movement patterns. We considered several possibilities: (1) developing semi-supervised models to divide the phone movement patterns into well-known human activities, (2) develop samples for human activities and apply sparse coding to identify the contexts (3) clustering the unlabeled behavioral footprints and use the cluster indices to train a Context Identification Model ($CIM$). However, we realized that it was not required to map the contexts to well-known human activities for developing the authentication systems, so we decided to implement (3).

\par
\begin{figure}[htp]
 \centering
 \includegraphics[width=3.35in, height=1.2in]{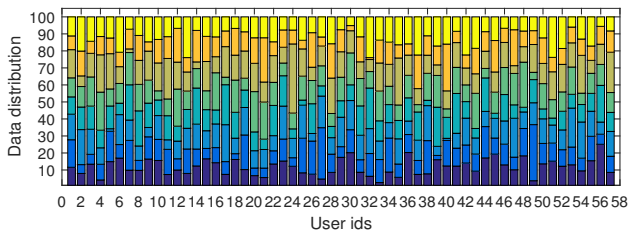}
 \caption{Illustrating the distribution of data across different clusters (contexts) for individual users. The same colors in bars do not mean the same type of context among the users. These contexts need not be mapped to any known human-activity to guide the authentication process. We simply numbered them so that we can identify and develop a separate authentication model for each of them.}
 \label{ContextWiseDataDistribution}
\end{figure}

 The next challenge was to define features that could help clustering algorithm in dividing the data not only based on the distance-based similarity but also on the structural-based similarity. This was one of the reasons to apply clustering of data at the feature-level not at the data-level. To capture the structural similarity of the signal as well as the pair of signals, we extracted a variety of features including spectral entropy, histograms (bins), Dynamic Time Warping (DTW) distance between the pair of signals ( acceleration in $X$, $y$, and $Z$ directions). The maximum number of clusters was fixed to eight in our experiment for each user. The distribution of data among the clusters varied drastically for every user (see Figure \ref{ContextWiseDataDistribution}). Some clusters had nominal samples, hence were neglected in the experiment. This behavior was expected as some users might have operated their phone in very different circumstances than the other ones. It would be interesting to find out the optimal number of clusters (contexts) that offers the best authentication accuracy.
\par A flowchart of the system is presented in Figure \ref{ExperiementalSetup}. The design of the authentication system could be divided into two phases: \textit{enrollment} and \textit{continuous verification}. The \textit{enrollment phase} in our system, included three steps: user-specific clustering of training data; training user-specific Context Identification Model ($CIM$) using the clusters and their indices; and finally training context-specific authentication models for each user. While the \textit{continuous verification phase} included identification of the context of the test samples using the $CIM$; selecting the appropriate authentication model using the context id produced by $CIM$; and then comparing the scores (assigned by the authentication model to the test samples) with a predefined threshold to make the authentication decision.
\subsection{Feature Extraction}
\label{FeatureExtraction}
The clustering, context classification, and user authentication were carried out at the feature level. The features were extracted from the signals formed by the accelerometer readings in the $x$, $y$, $z$, directions and their resultant $m$ which was defined as, $\sqrt(x^2+y^2+z^2)$. The features set consisted of descriptive statistical features such as mean, standard deviation, interquartile range, as well as features from frequency, spectral, and information theory domain e.g. band power, median frequency, spectral entropy, and mutual information, correlation, and Dynamic Time Warping distance between the pair of signals (see Table \ref{FeatureList}).
\par In order to implement the continuous authentication paradigm, we extracted these features by using sliding window protocol with ten seconds of window size and five seconds of overlapping \cite{Abena, TreadmiLL}. These numbers were derived from the existing body of the work in this domain. The screen on/off information was also used as a feature during both, the context classification and authentication process. The values of this feature were basically the percentage of times the screen was switched on in the windows of data used for feature extraction.
\subsection{Feature Analysis} To reduce the number of features, we evaluated all the extracted features by using the correlation based feature subset selection method with breadth first search method. The feature selection was performed separately for each context, for each user. The best of features that could distinguish between the genuine and the impostor classes for a particular context of a user, were selected and used to build the authentication model for that context of that user. On an average 70\% are more features were discarded through the correlation based feature selection method.
\par The scale of extracted features varied drastically, hence we normalized all the features between zero to one. To decide upon the normalization method, we tested all the features for normality using one-sample Kolmogorov-Smirnov test \cite{KSTest}. The null hypothesis of each test was that the feature values follow a standard normal distribution. The test resulted in one if it rejected the null hypothesis at the significance level of 5\%, or zero otherwise. The test results concluded that most of the features were not normally distributed, hence, we applied $min$\_$max$ normalization \cite{ScoreNormalization}.
\subsection{Genuine and Impostor Sample Considerations} The authentication models were implemented by using multi-class classifiers, hence, were trained by using samples from both genuine and impostors classes. For both, training and testing phases, the genuine samples were created using the data belonging to the actual (candidate) user, whereas the impostor samples were created using the data belonging to the rest of the users.
\begin{figure}[htp]
 \centering
 \includegraphics[width=3.4in, height=2.2in]{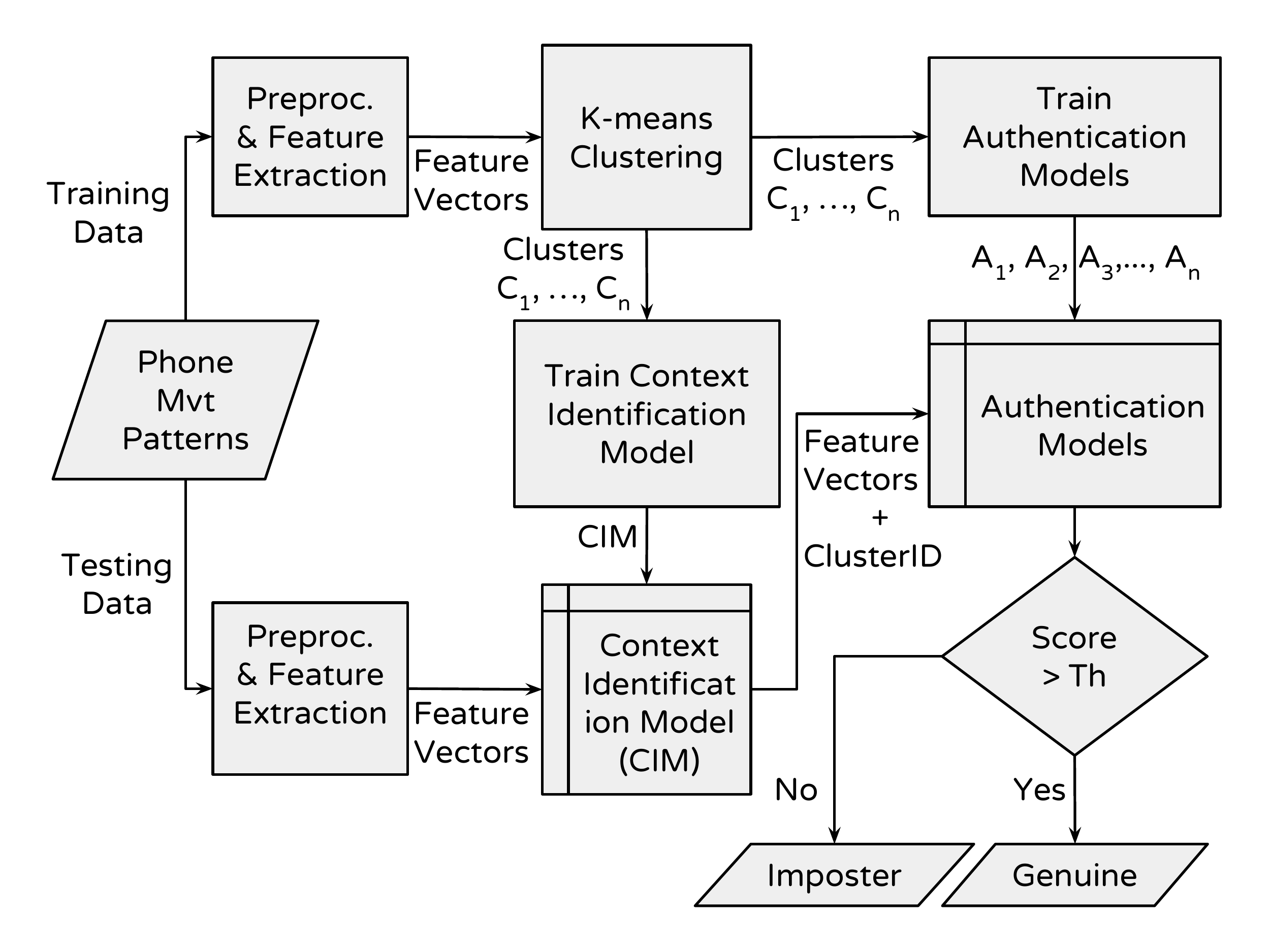}
 \caption{A flowchart illustrating the design of the system. The unlabeled phone movement patterns were equally divided into two parts training and testing. During training, the feature vectors went through the k-mean clustering that produced clusters. Assuming each cluster represented a context, context-specific authentication models were trained using five different classifiers. Similarly, $CIM$ was trained using Random Forest. During verification, the testing part of data was supplied to the preprocessor and the feature extractor to obtain feature vectors. The feature vectors were supplied to the Context Identification Model ($CIM$) which produced the context (or cluster) ID. The cluster ID was used to select the appropriate authentication model. The feature vectors were supplied to the selected authentication model to obtain the authentication decision.}
 \label{ExperiementalSetup}
\end{figure}
\subsection{Context Identification Model (CIM)}
\par The context identification model ($CIM$) was built separately for each user assuming that the users might have a different number of and distinct phone usage contexts. The number of context for each user indeed varied as shown in Figure \ref{ContextWiseDataDistribution}. To build the $CIM$, we clustered the normalized feature vectors using the k-means clustering algorithm. Assuming each cluster represented a distinct phone usage context, the cluster indices were assumed as class labels. To train $CIM$, we used feature vectors along with their corresponding cluster ID as class labels. The $CIM$ was implemented using the Random Forest classifier. Random Forest was chosen because it was proven to be effective in existing studies \cite{TreadmiLL, ArmMovement, ChungXu}.
\subsection{Identifying Phone Usage Contexts}
\par As shown in the Figure \ref{ExperiementalSetup}, testing samples were passed through $CIM$ of the corresponding user to find out the context of the test sample. Using the predicted context, the corresponding authentication model was selected that provided the authentication decision. For example, let the supplied test sample was classified into context $C_2$, then $A_2$ was used to make an authentication decision, if $A_2$ was build using the feature vectors belonging to cluster $C_2$ during enrollment.
\begin{figure*}[htp]
\centering
\begin{tabular}{cc}
\subfigure[Logistic Regression]{\epsfig{file=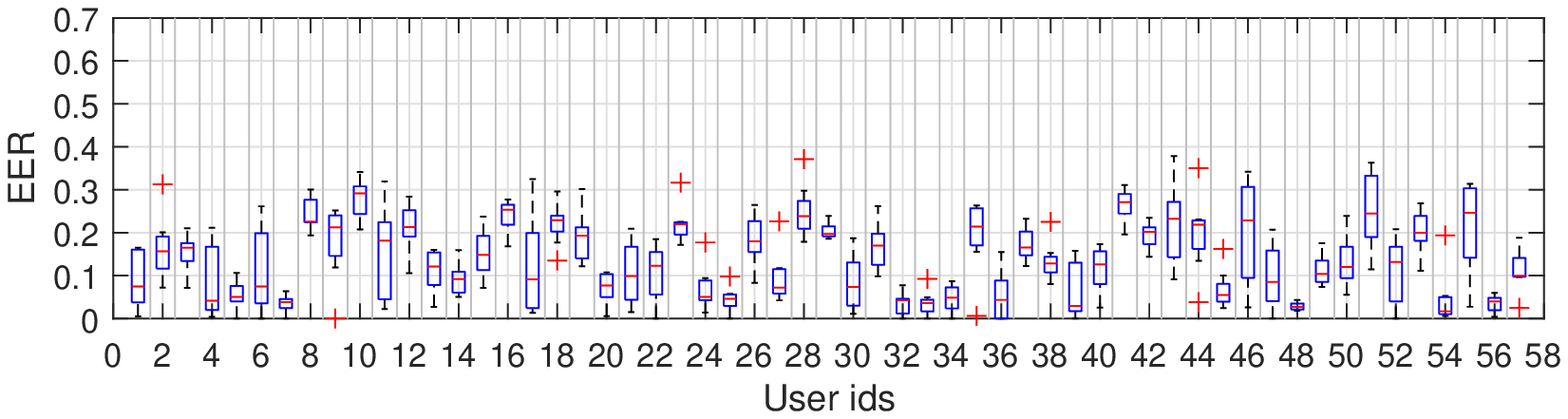,width=3.31in, height=1.3in}
\label{LogReg}}&
\subfigure[Neural Network]{\epsfig{file=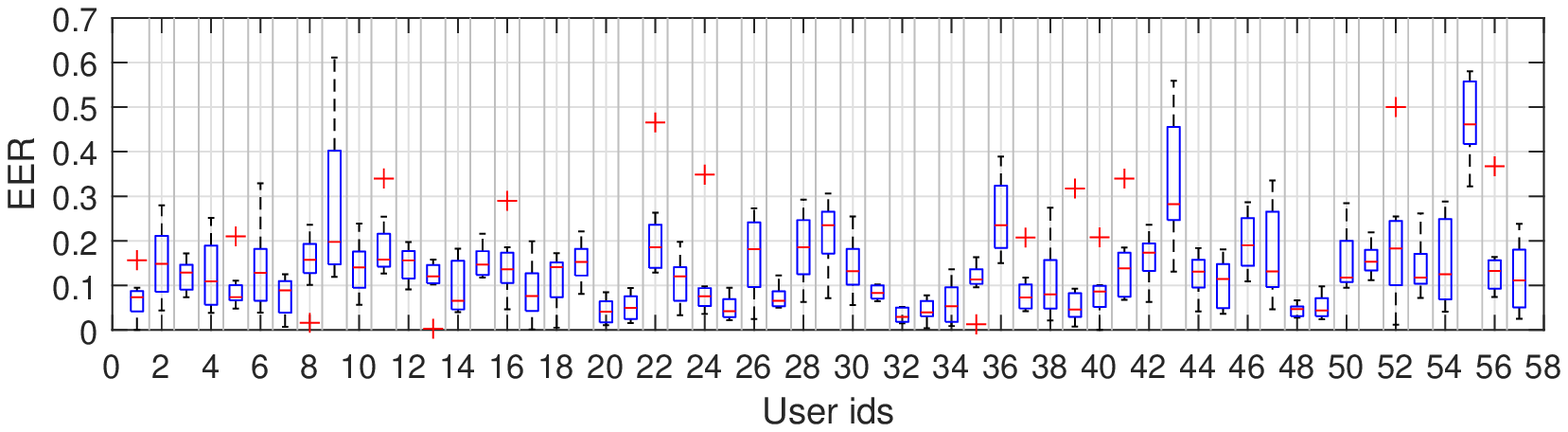,width=3.31in, height=1.3in}
\label{NNet}}\\
\subfigure[k Nearest Neighbors]{\epsfig{file=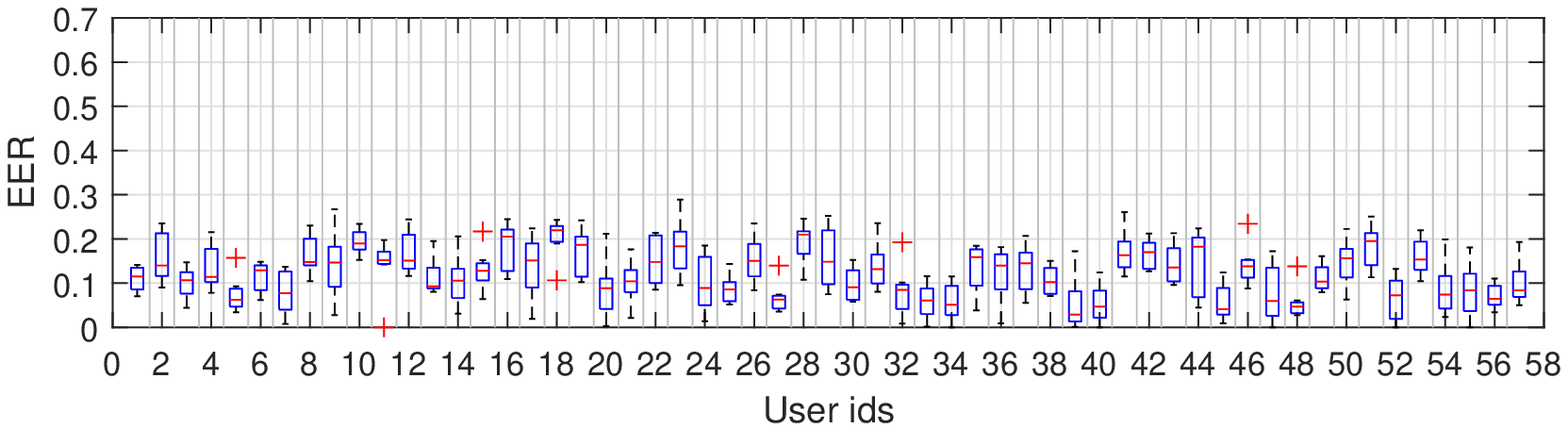,width=3.31in, height=1.3in}
\label{kNN}}&
\subfigure[Support Vector Machine]{\epsfig{file=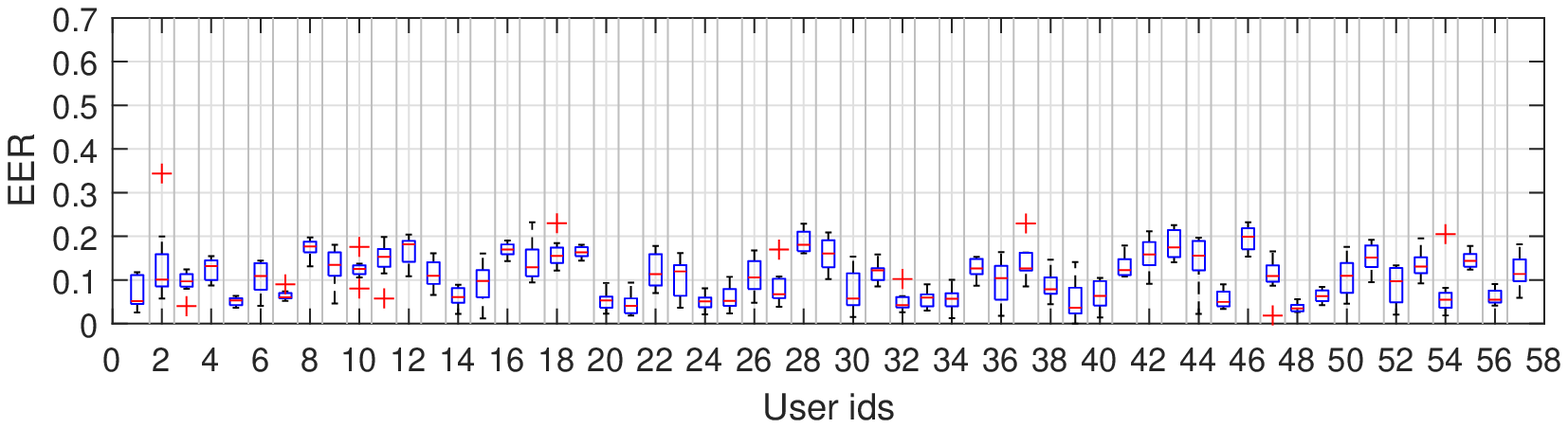,width=3.31in, height=1.3in}
\label{SVM}}\\
\subfigure[Random Forest]{\epsfig{file=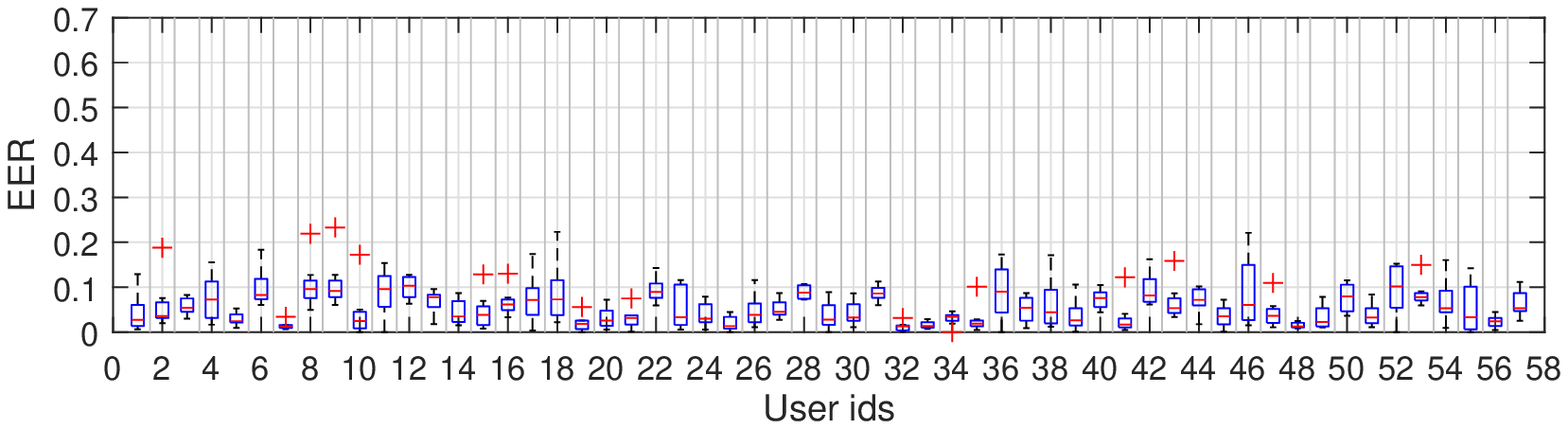,width=3.31in, height=1.3in}
\label{RandomForest}}&
\subfigure[Comparison of classifiers based on average EERs]{\epsfig{file=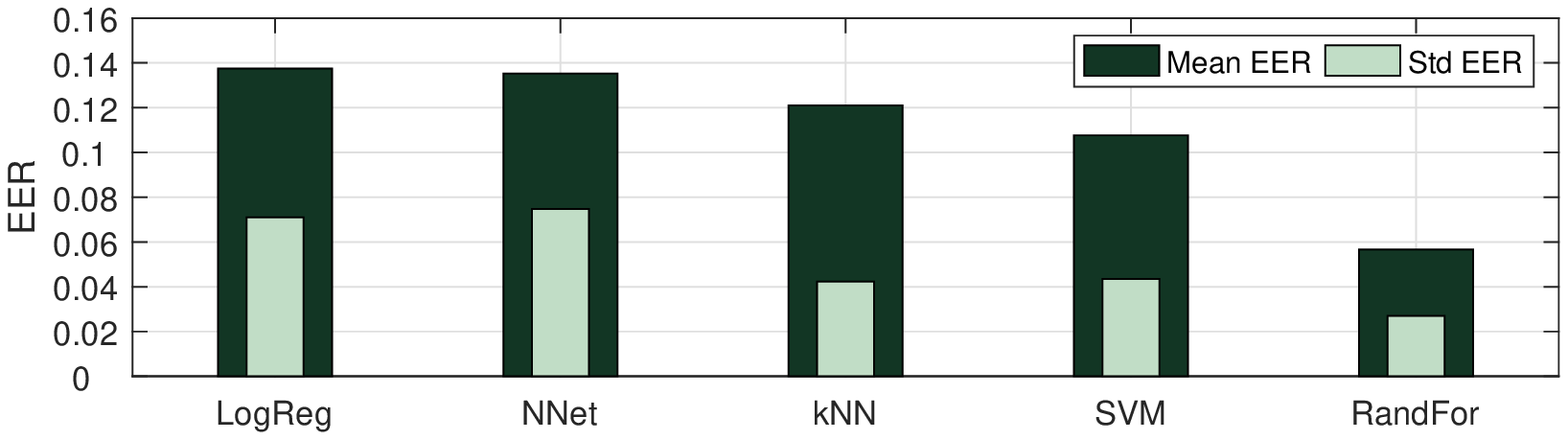,width=3.31in, height=1.25in}
\label{Comparision}}
\end{tabular}
\caption{Figures \ref{LogReg}, \ref{NNet}, \ref{kNN}, \ref{SVM}, and \ref{RandomForest} show the equal error rates achieved by Logistic Regression, Neural Network, k Nearest Neighbors, Support Vector Machine (SVM), and Random Forest respectively. While Figure \ref{Comparision} shows means and standard deviations of equal error rates achieved by the classifiers over all possible contexts of all the users. It is quite evident that the Random Forest wins the race in terms of the accuracy.}
\label{ClassPerformance}
\end{figure*}
\subsection{Context-wise Authentication Models}
\par Multiple authentication models, one for each context, were trained for each user. These classifiers needed knowledge of both genuine, and impostor classes. In our experimental setup, a fixed number of samples were borrowed from rest of the users and used as impostor samples. The ideal thing would be to choose any of the samples from other users as impostor samples than the candidate user. However, we used only those samples from other users as impostor samples that were classified (by $CIM$) in the similar context as of the candidate user.
\par The mapping of contexts among the users was challenging as the context $C_i$ of the $U_{candidate}$ may not necessarily match with the contexts $C_{i(s)}$ of $U_{impostors}$. Therefore, it would \textit{not} be ideal to use the samples from $C_i$ of user $U_{impostors}$ as impostor samples for training the authentication model $A_i$ for the context $C_i$ of the $U_{candidate}$. To address this problem, we simply used the $CIM$ of $U_{candidate}$ that classified samples of $U_{impostors}$ into one of the contexts of $U_{candidate}$. For example, to build the authentication model $A_2$ for context $C_2$ of user $U_{1}$, the samples belonging to $C_2$ of user $U_{1}$ were used as genuine samples. While samples of other users $U_2$ - $U_{57}$ were first classified using the $CIM$ of $U_{1}$, then, a portion of those samples that were classified as $C_2$ was used as impostor samples for training $A_2$.
\subsection{Verification and Performance Evaluation}
Five different classifiers, namely, Logistic Regression, Neural Network, k nearest neighbor (kNN), support vector machine (SVM), and Random Forest were used for implementing the authentication models. One of the reasons to employ these classifiers was their effectiveness in the existing studies \cite{Abena, TreadmiLL, PhoneGait1, ChungXu, HMOG, ArmMovement, TypingSwipingFusion, RoboticRobbery}. Also, their suitability for solving both linear and nonlinear recognition problems as well as have distinct operational characteristics. The setting for Logistic Regression classifiers was the generalized model with binomial family. The number of neurons in the hidden layer of Neural Network (Multilayer Perceptron) was set to 10. The k-NN was implemented using the k=10. The SVM was used with RBF Kernel and C-classification settings. Rest of the settings for all classifiers were left to default as provided by respective packages of R language.
\par All authentication models were tested for both genuine and impostor pass rates. The prediction probabilities for unknown samples that came from $U_{genuine}$ were referred to as the genuine score. While the prediction probabilities for unknown samples that came from $U_{impostors}$ were referred to as the impostor scores. The performance of every authentication model was evaluated using the equal error rate (EER). The error rate obtained for a threshold at which false accept and false reject rates match is known as the EER. The EER was computed using the genuine scores, impostor score, and varying thresholds. The EER has been extensively used to compare the authentication systems. The lower the EER is the better the authentication system.
\section{Experimental Results and Discussion}
\label{ExperimentalResultsAndDiscussion}
\subsection{Performance Across User Population}
The results of all experiments are summarized in Figure \ref{ClassPerformance}. All sub figures except the one in the bottom-right corner represent the user-wise performance of Logistic Regression, Neural Network, kNN, SVM, and Random Forest respectively. The box plot for each user was plotted using the mean EERs for each context. The box plots depict the median EER (center red line) for each user. Red cross whiskers represent the EERs of the outliers contexts. Figure \ref{Comparision} illustrates the average and standard deviation of EERs computed across all the contexts of all 57 users. The average equal error rate were 13.7\%, 13.5\%, 12.1\%, 10.7\%, and 5.6\% with the standard deviation of 7\%,  7\%, 4.2\%,  4.3\%, 2.7\% for Logistic Regression, Neural Network, kNN, SVM, and Random Forest classifiers respectively. The median EERs achieved by Random Forest, SVM, and kNN were very consistent across the user population. While the Neural Network has performed very well for most of the users but quite poorly for few users. A similar trend could be observed for the Logistic Regression classifiers.
\begin{table}[htp]
\centering
\scriptsize
\caption{Results of statistical test for significance of difference in the mean EERs obtained by the different machine learning classifiers.}
\vspace{0.1in}
\begin{tabular}{|c|c|c|c|}
\hline
\multirow{3}{*}{\begin{tabular}[c]{@{}c@{}}Pair of \\ Algorithms\end{tabular}} & \multicolumn{3}{c|}{p-values}                                       \\ \cline{2-4}
                                                                           &Normality & \multicolumn{2}{c|}{Performance} \\ \cline{2-4}
                                                                           &KS       &Friedman              &Wilcoxon \\ \hline\hline
Random Forest - SVM                                                                &4.63E-13       &1.43E-11                   &8.74E-11             \\ \hline
Random Forest - Neural Network                                                               &3.57E-13       &3.22E-13                   &5.43E-11             \\ \hline
Random Forest - Logistic Regression                                                             &1.80E-12       &2.22E-12                   &2.24E-10             \\ \hline
Random Forest - kNN                                                                &1.21E-12       &2.22E-12                   &1.02E-10             \\ \hline
SVM-Neural Network                                                                   &4.16E-12       &1.18E-02                   &0.0025               \\ \hline
SVM-Logistic Regression                                                                 &2.32E-12       &3.49E-04                   &6.52E-06             \\ \hline
SVM-kNN                                                                    &5.67E-12       &3.49E-04                   &2.17E-04             \\ \hline
kNN-Neural Network                                                                  &1.92E-11       &0.6911                     &0.6421               \\ \hline
kNN-Logistic Regression                                                                 &7.43E-12       &0.0469                     &0.0088               \\ \hline
Neural Network-Logistic Regression                                                                &1.54E-10       &0.8946                     &0.5328               \\ \hline
\end{tabular}
\label{StatsTest}
\end{table}
\begin{figure*}[htp]
\centering
\begin{tabular}{cccc}
\subfigure[All 57 users]{\epsfig{file=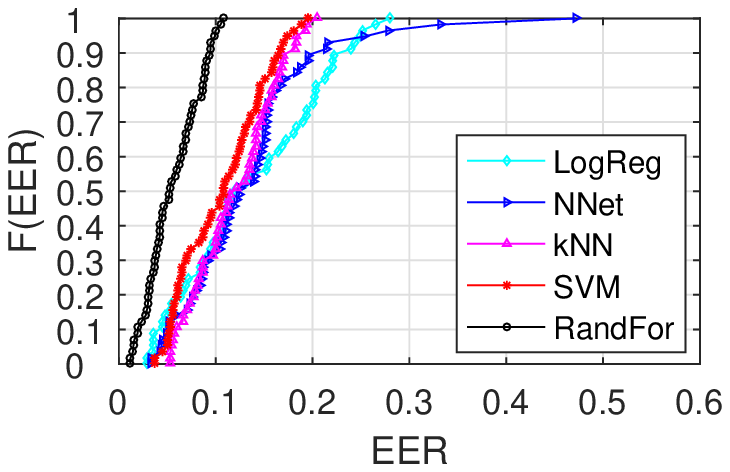,width=1.66in, height=1.24in}
\label{Users100}}
\subfigure[After removing 5\% bad users]{\epsfig{file=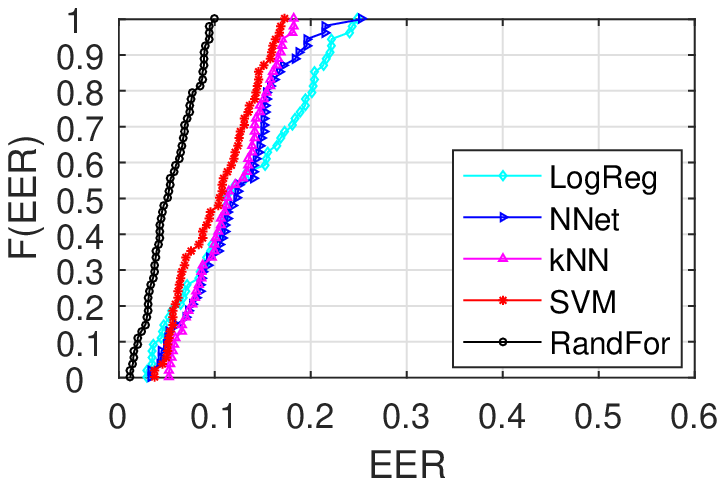,width=1.66in, height=1.24in}
\label{Users95}}
\subfigure[After removing 10\% bad users]{\epsfig{file=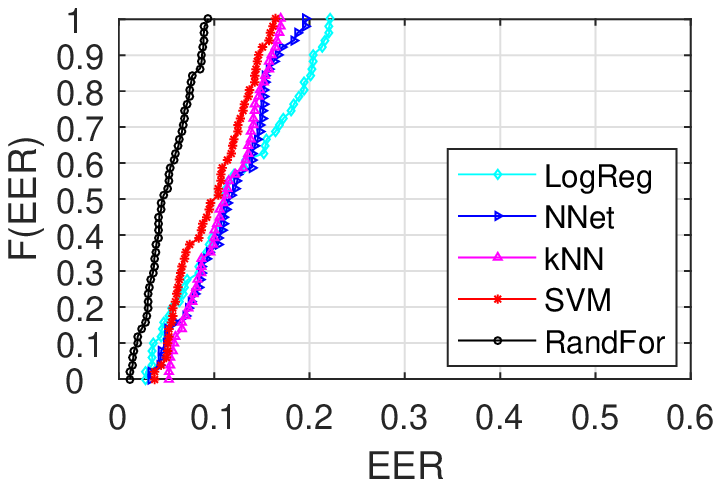,width=1.66in, height=1.24in}
\label{Users90}}
\subfigure[After removing 15\% bad users]{\epsfig{file=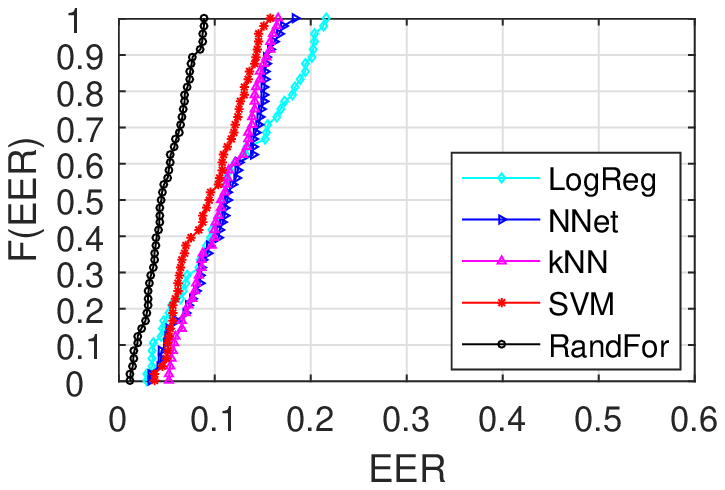,width=1.66in, height=1.24in}
\label{Users85}}
\end{tabular}
\caption{The y-axis represents the cumulative distribution mean EERs obtained by all five classifiers. Figures \ref{Users100}, \ref{Users95}, \ref{Users90}, and \ref{Users85} show the cumulative distribution of mean EERs after the removal of 0, 3, 6, and 9 bad users systematically.}
\label{FailureToEnroll}
\end{figure*}
\subsection{Comparison of the Classifier's Performance} The average and standard deviations of EERs could be misleading while ranking the quality of these classifiers for authentication purpose. Figure \ref{Comparision} gives an impression that the Neural Network-based authentication systems exhibited significantly inferior performance compared to kNN, and SVM based systems. However, that may not be the case if we remove a few users that exhibited outlying error rates. Hence, to measure the quality of these classifiers, a series of statistical tests was conducted.
\par The mixed effects Analysis of Variance (ANOVA) is generally used to test the statistical significance of differences between mean error rates of the two different methods. Mixed effects ANOVA works under the assumption that the underlying distribution of the pairwise difference of error rates follows a Gaussian distribution \cite{Friedman}. Therefore, first, we tested the pairwise difference of mean EERs for the normality using Kolmogorov-Smirnov (KS) test \cite{KSTest}. The null hypothesis was that the pairwise difference of mean EERs obtained by two algorithms follows a Gaussian distribution. KS test rejected the null hypothesis at the 5\% significance level for all pair of algorithms. Since for all pair of algorithms, the pair-wise difference between mean error rates failed to pass the normality test, we chose to use Friedman test \cite{Friedman}.
\par The Friedman test was carried out under the null hypothesis that the algorithms are equally accurate. Table \ref{StatsTest} shows that the test resulted in significantly low p-value (less than $10^{-2}$), for all pairs of the classifiers except for (kNN, NNet), (kNN, LogReg), and (NNet, LogReg). Therefore we failed to reject the null hypothesis for kNN, NNet, and LogReg at the significance level of 3\%. We also performed the Wilcoxon signed rank test. The test results affirmed the same conclusion as of Friedman (see Table \ref{StatsTest}). As the statistical tests demonstrated that the mean EERs obtained by kNN, NNet, and LogReg are sufficiently similar, we conclude that differences among the performances of kNN, NNet, and LogReg are statistically insignificant.
\subsection{User-wise Suitability of the Proposed Systems}
The behavioral footprints of individuals are subjected to change under different operating conditions. Our proposed model addresses that concern up to a certain extent by clustering the patterns. The same behavioral trait, if significantly different under different conditions, could get clustered under two or more contexts depending on what operating conditions that behavioral activity occurred. In Figure \ref{ClassPerformance}, we could see that few users have significantly high error rates (e.g. user ids $9, 36, 43,$ and $55$). We referred to the users who have high error rates as bad users. The high error rates could be because of the following reasons: (1) the bad users have high variance in their own data, and/or (2) their phone movement pattern overlap too much with the other users.
\par Through Figure \ref{FailureToEnroll}, we could see the impact of the systematic removal of bad users on the overall performance of the proposed systems. The removal of top 3 (5\% of total) bad users caused the Random Forest consistently maintain the mean EERs below 10\% for each user. SVM, kNN, and Neural Network could achieve below 10\% mean EERs for half of the users, while the rest of the users fell under 10-20\%. However, Logistic Regression still had more than 20\% of the total users who had beyond 20\% EERs, and more than 40\% users had EERs between 10-20\%. Likewise, removal of top 9 (15\% of total) bad users brought SVM, kNN, and Neural Network in the same league in terms of mean EERs. Needless to say that the Random Forest was the most suitable algorithm, whereas, Logistic Regression was the least.
\vspace{-0.1in}
\section{Conclusion and Future Work}
\label{ConclusionAndFutureWork}
We conclude that it is possible to continuously authenticate smartphone users using only phone movement patterns. Our experimental results hold Random Forest as the most suitable classifier for implementing such authentication system. The Random Forest outperformed SVM, kNN, Neural Network, and Logistic Regression. User-wise suitability of the proposed system was also investigated. The results suggested that the proposed system may not be suitable for certain types users and could exhibited quite high error rates. The exclusion of those users, however, could improve the overall performance significantly. Therefore, we conclude that the phone movement pattern based authentication systems may not be suitable for every smartphone user. In the future, we plan to explore: the suitability of one class classifiers for the above system as decision on selecting the impostor samples for training was difficult; explore the fusion of phone movement patterns with other biometric modalities in order to improve the authentication accuracy; and the investigate the fusion of decision of authentication models based on different classification algorithms.
\section{Acknowledgement} This work was supported in part by DARPA Active Authentication contract FA \#8750-13-2-0274 and by National Science Foundation Award SaTC \#1527795.
{\small
\bibliographystyle{ieee}
\bibliography{submission_example}
}

\end{document}